\documentclass[aps,prl,twocolumn,superscriptaddress,longbibliography]{revtex4-1}
\usepackage{graphicx}
\usepackage{amssymb,amsmath}
\usepackage{bm}
\usepackage{dcolumn}
\usepackage{subfigure}
\usepackage{float}
\usepackage[OT1]{fontenc} 
\usepackage{url}
\usepackage{slashed}
\usepackage{color}
\usepackage{verbatim}
\usepackage{times}
\usepackage{graphicx}
\usepackage[colorlinks,linkcolor=blue,anchorcolor=blue,citecolor=blue]{hyperref}
\usepackage{ dsfont }

\begin{document}
\pdfoutput=1
\title{ Preparing Quantum States by Measurement-feedback Control with Bayesian Optimization}

\author{Yadong Wu}
\affiliation{Department of Physics, Fudan University, Shanghai, 200438, China}
\affiliation{State Key Laboratory of Surface Physics, Key Laboratory of Micro and Nano Photonic Structures (MOE), Fudan University, Shanghai 200438, China}
\affiliation{Institute for Nanoelectronic Devices and Quantum Computing, Fudan University, Shanghai 200433, China}
\affiliation{Shanghai Qi Zhi Institute, AI Tower, Xuhui District, Shanghai 200232, China}

\author{Juan Yao}
\affiliation{Shenzhen Institute for Quantum Science and Engineering, Southern University of Science and Technology, Shenzhen 518055, Guangdong, China}
\affiliation{International Quantum Academy, Shenzhen 518048, Guangdong, China}
\affiliation{Guangdong Provincial Key Laboratory ofQuantum Science and Engineering, Southern University of Science and Technology, Shenzhen 518055, Guangdong, China}

\author{Pengfei Zhang}
\email{pengfeizhang.physics@gmail.com}
\affiliation{Department of Physics, Fudan University, Shanghai, 200438, China}
\affiliation{Walter Burke Institute for Theoretical Physics \& Institute for Quantum Information and Matter, California Institute of Technology, Pasadena, CA 91125, USA}

\begin{abstract}

Preparation of quantum states is of vital importance for performing quantum computations and quantum simulations. In this work, we propose a general framework for preparing ground states of many-body systems by combining the measurement-feedback control process (MFCP) and the machine learning method.  Using the Bayesian optimization (BO) strategy, the efficiency of determining the measurement and feedback operators in the MFCP is demonstrated. Taking the one dimensional Bose-Hubbard model as an example, we show that BO can generate optimal parameters, although constrained by the operator basis, which can drive the system to the low energy state with high probability in typical quantum trajectories.

\end{abstract}
	
\maketitle
\textbf{Introduction}. 
Quantum states preparation is an essential part of the research in the quantum physics field. Recently, it has been realized that measurements play an important role in preparing quantum states \cite{geremia2004real,PhysRevLett.97.073601,PhysRevLett.97.190201,PhysRevLett.99.223601,Sayrin:2011uk,PhysRevLett.108.243602,Riste:2013um,PhysRevLett.110.163602,PhysRevLett.115.060401,PhysRevLett.116.093602,PhysRevLett.117.073604,PhysRevA.94.052120,PhysRevX.7.011001}
.  
For states that can be specified by a set of stabilizers, including both the Greenberger-HorneZeilinger `cat' state and the toric code state, efficiently preparation algorithm have been proposed by measuring cluster states and applying local unitary gates based on the measurement outcome \cite{briegel2001persistent,raussendorf2005long,verresen2021efficiently,tantivasadakarn2021long,zhu2022nishimori,lee2022measurement,tantivasadakarn2022hierarchy,bravyi2022adaptive,lu2022measurement}. This is an example of the measurement feedback control process (MFCP), where we measure the quantum system during the evolution of a quantum trajectory, obtain the measurement signal with random noise, and feedback controls the quantum system based on measurement outcomes \cite{PhysRevA.49.2133, zhang2017quantum}.  MFCP is a valuable approach to affect the quantum system in a partially human-controlled way and has been wildly adopted in cold atoms, ion traps, photon cavities, and other systems \cite{PhysRevA.65.061801,PhysRevLett.111.020501,Grimsmo_2014,Kopylov_2015,Mazzucchi:16,PhysRevLett.122.233602,Ivanov:2020tf,PhysRevLett.124.010603,Kroeger_2020,PhysRevLett.124.110503,PhysRevA.102.022610,PhysRevResearch.2.043325,PhysRevA.104.033719}. 

Motivated by these developments, we explore the possibility of preparing more general ground states of lattice Hamiltonians using MFCP. We focus on MFCP with continuous weak measurements, for which concrete realization has been proposed in cold atom systems \cite{PhysRevA.99.053612,PhysRevResearch.3.043075}. Since the measurement signal is a random variable with a large variance because of the uncertainty principle, we get an ensemble of quantum trajectories instead of a single quantum state \cite{wiseman2009quantum}. The protocol succeed only if typical quantum trajectories, instead of the averaged density matrix, have high fidelity with respect to the target. However, the MFCP contains much arbitrariness: Generally, it is not clear how to choose suitable measurement and feedback control operators to achieve the target state.

\begin{figure}
	\includegraphics[width=.98\columnwidth]{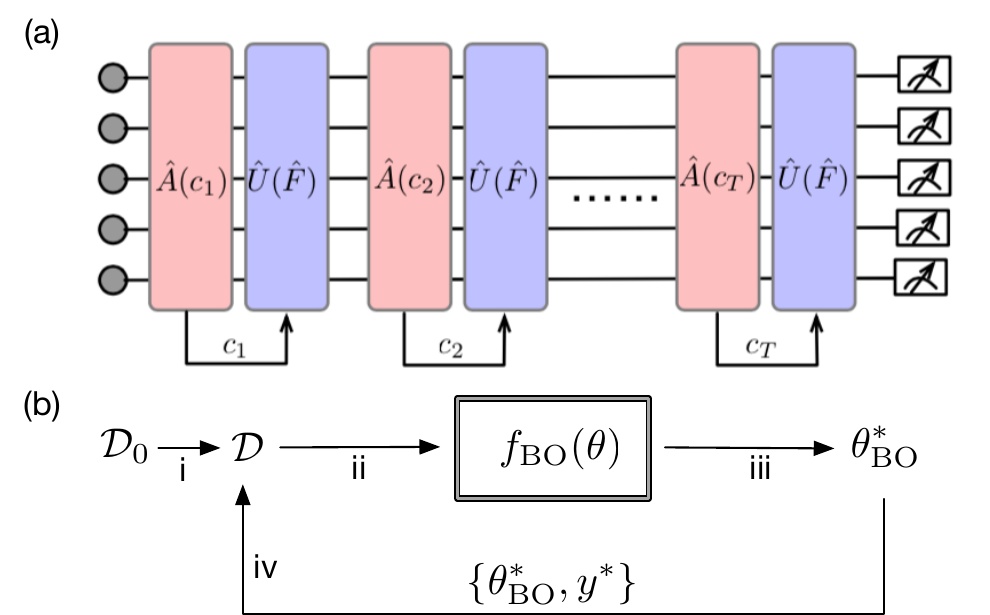}
	\caption{(a) MFCP for a quantum system. $\hat{A}(c_m)$ is the measurement step and $\hat{U}(\hat{F})$ is the feedback control step. At each time interval, the feedback control step evolves the system with the measurement signal $c_m$. After $T$ times the system is detected. (b) The general procedure of the BO method to obtain the global extremum of a 'black-box' function $f(\theta)$.} 
	\label{flow}
\end{figure} 

 In this work, we utilize the machine learning (ML) method to overcome this difficulty and achieve high fidelity of preparing target states. There are many beautiful works on using ML in physics, including classifying different quantum phases, accelerating computation, preparing physical systems, and optimizing experimental control processes \cite{PhysRevB.97.205140,wang2017machine,PhysRevLett.120.066401,wang2018machine,sun2018deep,PhysRevB.100.045153,wang2019emergent,Zhang:2019ui,Rem:2019vc,Bohrdt:2019tr,yao2020active,PhysRevLett.123.230504,Palmieri:2020vc,YadongWu2020,saggio2021experimental,baum2021experimental,castaldo2021quantum,sivak2022model,erdman2022identifying,langer2022representations,luchnikov2022probing,khait2022optimal,Carrasquilla:2017wu,Nieuwenburg:2017tz,PhysRevLett.118.216401,PhysRevB.96.195145,PhysRevLett.120.176401,PhysRevB.99.121104,carleo2017solving,Gao:2017uk,PhysRevB.97.035116,doi:10.7566/JPSJ.87.074002,Torlai:2018wn,wu2018visualizing,wang2020machine}. Among different ML algorithms, Bayesian optimization (BO) \cite{shahriari2015taking} stands out for obtaining the global maximum or minimum of a 'black-box' function where the function form is complicated but unknown. It's usually applied in a situation where the label of the function is expensive to obtain or the dataset is not abundant, such as parameters control and experimental calibration \cite{vargas2019bayesian,mukherjee2020preparation,kuros2020phase,sauvage2020optimal,chu2021bayesian,kuros2021controlled,xie2022bayesian,cortes2022sample}. We take the one-dimensional Bose-Hubbard model as an example and use BO to find the suitable measurement and feedback control operators for the ground-state preparation. After training, the MFCP can evolve the quantum system to states with low energy expectations, which are quite close to the true ground state.

\vspace{5pt}

\textbf{Method}. 
\textit{Measurement feedback control process.}
For systems evolved under the MFCP, we can divide the evolution time into a sequence of the interval length $\delta t$. At each time step, the evolution is composited of a weak measurement and an unitary evolution with feedback control Fig. \ref{flow}(a). The weak measurement is described by a set of Kraus operators with $\{\hat A(\alpha)\}$ with $\alpha\in \mathds{R}$, which are chosen to be a Gaussian function of a Hermitian measurement operator $\hat{c}$ \cite{jacobs2006straightforward}:
\begin{align}
\hat{A}(\alpha)&=\left(\frac{4\gamma\delta t}{\pi}\right)^{1/4} e^{-2\gamma\delta t(\hat{c}-\alpha)^2}.
\end{align}
Here $\gamma$ is the dissipative strength and $\alpha$ is the measurement result. The normalization factor is chosen to satisfy $\int^\infty_{-\infty}d\alpha ~ \hat{A}(\alpha)^{\dagger}\hat{A}(\alpha)=\hat{I}$. Given a state $|\psi\rangle$, during each time interval, measurement result $c_m$ satisfies the normal distribution $P(c_m)=\langle\hat{A}(c_m)^\dagger \hat{A}(c_m)\rangle\sim \mathcal{N}(\mu=\langle c\rangle, \sigma^2=1/8\gamma\delta t)$. Introducing the measurement signal $\delta W$, the measurement result can be rewritten as $c_m= \langle\hat{c}\rangle+\frac{\delta W}{\sqrt{8\gamma}\delta t}$, where the measurement signal $\delta W\sim \mathcal{N}(0,\delta t)$ represents the random noises in each weak measurement. After each weak measurement, the normalized quantum state evolves according to
\begin{align}
\label{Measure}
\mathcal{M}_t|\psi(t)\rangle =\frac{1}{\mathbf{N}} e^{-2\gamma(\hat{c}-\langle \hat{c}\rangle)^2\delta t+\sqrt{2\gamma}(\hat{c}-\langle\hat{c}\rangle)\delta W}|\psi(t)\rangle ,
\end{align}
which takes the form of an imaginary-time evolution with self-consistent field $\langle\hat{c}\rangle$ and random field $\delta W$. Here $1/\mathbf{N}$ is the normalizing factor.

A feedback step is followed in order to control the system. We applying the Markovian feedback control to the system with a feedback term proportional to the measurement signal $c_m$ added to the system. The unitary dynamic is described by
\begin{align}
\label{Feedback}
U_t|\psi(t)\rangle=e^{-i(\hat{H}_0+c_m\hat{F})\delta t/\hbar}|\psi(t)\rangle
\end{align}
where $\hat{F}$ is the feedback operator and $\hat{H}_0$ is the system Hamiltonian. Weak measurement (Eq.\eqref{Measure}) and feedback step (Eq.\eqref{Feedback}) together contribute one loop of the dynamic evolution:
\begin{equation}
|\psi(t+\delta t)\rangle=U_t\mathcal{M}_t|\psi(t)\rangle.
\end{equation}
As shown in Fig.\ref{flow}(a), repeating multiple times of the measurement-dependent feedback loop, the system is evolved to $|\psi(T)\rangle$.

The measurement-dependent feedback control can be used to tailor the system's dynamics and states. Proper choice of measurement operator $\hat{c}$ and feedback operator $\hat{F}$ can lead to high fidelity between the evolved state $|\psi(T)\rangle$ and some target state such as the ground state $|\psi_g\rangle$ of some Hamiltonian. In most cases, the explicit form of the ground state is unknown and fidelity between the two states is unavailable. Alternative target functions such as state energy or entropy can be used to guide the choice of $\hat{c}$ and $\hat{F}$. Thus the key problem lies in how to determine the form of $\hat{c}$ and $\hat{F}$ which optimize the target function.  In this work, we choose some operator basis and parameterize both $\hat{c}$ and $\hat{F}$ by their expansion coefficients $\alpha_j$ and $\beta_j$.  In the following, we will adopt the machine learning method to determine the optimized parameters $\theta=\{\alpha_j,\beta_j\}$ for this measurement-dependent feedback control problem.

\textit{Bayesian optimization.}
Bayesian optimization (BO) is a strategy for global optimization. The ultimate goal of the BO method is to search the extremum point of a black-box function $f(\theta)$. In order to simulate the relation between variables $\theta$ and $y=f(\theta)$, sampling of training dataset with $\mathcal{D}=\{(\theta_i,y_i)|i=1,\cdots\}$ are required. In some cases, the label $y$ of the function is hard to obtain. For example, it may require conducting a time consuming experiment for each parameter $\theta$. BO method queries the labels online in an economical way and extends the training dataset through a small training dataset $\mathcal{D}_0$. It can obtain the global extremum with a limited number of labels which saves an amount of time and resources. 

The general procedure of the BO method is summarized in Fig. \ref{flow}(b). i) Initially, a small training dataset $\mathcal{D}=\mathcal{D}_0=\{(\theta_i,y_i)|i=1,\cdots,n_0)\}$ is prepared. ii) Training the BO model with the dataset $\mathcal{D}$ and obtaining the approximated model $f_{\text{BO}}(\theta)$. iii) Searching the extremum point $\theta^*_{\text{BO}}$ of the training BO model $f_{\text{BO}}(\theta)$. iv) Querying the true label $y*=f(\theta^*_{\text{BO}})$ and updating the training dataset $\mathcal{D}$ by adding the new sample point $(\theta^*_{\text{BO}}, y^*)$. Repeating steps ii)-iv) until the prediction is converged. Then we can claim that the BO optimization approaches the extremum point of the function $y^*=f(\theta^*)$. Although the training dataset is extended during the training process, it is possible that the limited number of training data points is not large enough to simulate the real function $f(\theta)$ in entire regimes. Nevertheless, the simulated $f_{\text{BO}}$ is accurate enough to predict the extremum point of the $f(\theta)$ due to the query strategy adopted by the BO method. 

In our setup, we are mainly interested in preparing ground states for lattice Hamiltonians. Comparing to using fidelity as the target function, the energy is more realistic to be measured in experiments. As a result, we set the target function as the energy expectation values with $f(\theta)\equiv \overline{\langle\psi(T)|\hat{H}|\psi(T)\rangle}$. For states with long-range orders at zero temperature, the correlation length of order parameters may also be used to be the target function. Due to the stochastic property of the measurement result and the associated feedback control process, for each set of $\theta$, each running of the dynamic evolution is only one of the many possible quantum trajectories. Considering the real experimental operation condition, we repeat the dynamic evolution 50 times and only take the first 25 quantum trajectories with minimal energy expectation values. Then for each set of parameters $\theta$, $\overline{(\cdot)}$ corresponds to the average of the 25 quantum trajectories. All training processes are implemented in MATLAB Bayesian optimization toolbox.

\textit{Analytical Benchmark.}
Before turning to the example, we provide an analytical benchmark for our BO based protocol.
Up to the fisrt order of $\delta t$, the MFCP is reduced to the stochastic master equation \cite{zhang2017quantum},
\begin{align}
\label{SME}
d\hat{\rho} &= -i[\hat{H}_0+\hat{H}_{\text{fb}},\hat{\rho}]dt+\mathcal{D}[\hat{L}]\hat{\rho}dt+\mathcal{H}[\hat{L}]\hat{\rho}dW,
\end{align}
where $\mathcal{D}[\hat{L}]\hat{\rho} = \hat{L}\hat{\rho}\hat{L}^\dagger-\frac{1}{2}(\hat{L}^\dagger\hat{L}\hat{\rho}+\hat{\rho}\hat{L}^\dagger\hat{L}),\mathcal{H}[\hat{L}]\hat{\rho}= \hat{L}\hat{\rho}+\hat{\rho}\hat{L}^\dagger-{\rm Tr}[(\hat{L}+\hat{L}^\dagger)\hat{\rho}]\hat{\rho}$. $\hat{L}=(\sqrt{\gamma}\hat{c}-i\hat{F})$ is the modified dissipation operator and $\hat{H}_{\text{fb}}=\sqrt{\gamma}(\hat{c}^\dagger\hat{F}+\hat{F}\hat{c})$ is the modified feedback control Hamiltonian. After taking the ensemble average of Eq. (\ref{SME}) with $E[dW]=0$, the reduced master equation has a pure steady state which is the common eigenstate of $\hat{L}$ and $\hat{H}_{\text{eff}}=\hat{H}_0+\hat{H}_{\text{fb}}-i\hat{L}^\dagger\hat{L}/2$. As a result, by requiring $\hat{L}|\psi_{g}\rangle=0$, it is possible to reach a steady state very close to the ground state $|\psi_g\rangle$ of $\hat{H}_0$ when the feedback control Hamiltonian $\hat{H}_{\text{fb}}$ can be neglected for the weak measurement with strength $\gamma\sim0$. Provided by the ground state $|\psi_g\rangle$, it is efficient to minimize the expectation value of $\langle\hat{L}^\dagger\hat{L}\rangle=\text{Tr}[\hat{L}^\dagger\hat{L}|\psi_g\rangle\langle\psi_g|]/\text{Tr}[\hat{L}^\dagger\hat{L}]$. This minimization procedure is equivalent to search the eigenvector with minimal eigenvalue of a semi-positive matrix $M$. Here the element of the matrix $\mathcal{K}_{lm}=\langle \psi_g | \hat{O}_l^\dagger \hat{O}_m|\psi_g\rangle$ with $\hat{O}_i$ composing the operator basis for parametrization of $\hat{L}$, which is known as a correlation matrix \cite{qi2019determining,bairey2019learning,chertkov2018computational,li2020hamiltonian,PhysRevResearch.4.L042037}. The corresponding eigenvector with the minimal eigenvalue provides the expansion coefficients $\theta_{\rm L}$ with associated $\hat{L}_{\mathcal{K}}$, which can drive the system close to the ground state of $\hat{H}_0$.  For the example discussed in the following, the ground state $|\psi_g\rangle$ is obtained numerically with exact diagonalization method.  

\vspace{5pt}

\textbf{Example}. 
Now we present an example of our protocol using the Bose-Hubbard model. MFCP has been applied to prepare both the ground state and excited states in \cite{wu2022cooling}, where measurement and feedback operators are constructed based on the perturbative analysis in the non-interacting case. As a consequence, the performance decreases as the interaction strength increases. In the following, we take the Bose-Hubbard model as an example to illustrate how to determine the measurement and feedback operator with the BO method. Considering $N$ bosonic particles in a one-dimensional optical lattice of $M$ sites.  The Hamiltonian reads 
\begin{align}
\hat{H}_0=-J\sum_j^{M-1}\hat{a}^\dagger_{j}\hat{a}_{j+1}+\hat{a}^\dagger_{j+1}\hat{a}_j+\frac{U}{2}\sum_j^M\hat{n}_j(\hat{n}_j-1)
\label{Hami}
\end{align}
where $J$ is the nearest-neighbor hopping strength, $U$ is the contact interaction strength. Here $\hat{a}_j$ is the bosonic annihilation operator at the $j_{th}$ site and $\hat{n}_j=\hat{a}_j^\dagger\hat{a}_j$ is the particle number operator. For experimental convenience, the measurement operator is set as the linear combination of the single-site particle number operators 
\begin{align}
\hat{c}=\sum_j^M\alpha_j\hat{n}_j,
\end{align}
and the feedback operator is proportional to the nearest neighbor hopping with
\begin{align}
\hat{F}=(\beta_1+i\beta_2)\sum_j^{M-1}\hat{a}_j^\dagger\hat{a}_{j+1}+\text{H.C.},
\end{align}
where an overall modification of the hopping amplitude and phase is applied to the system. Realization of the scheme for the measurement operator $\hat{c}$ has been discussed in \cite{RevModPhys.85.553,PhysRevLett.114.113604,PhysRevLett.115.095301}. The feedback operator $\hat{F}$ can also be implemented by using photon-assisted hopping \cite{zhang2017quantum}. Then the coefficients $\theta = \{\alpha_j$, $\beta_1$, $\beta_2|j=1,...,M\}$ will determine the final state of MFCP $|\psi(T)\rangle$. The same operator basis will also be used  for the analytic benchmark analysis.

\begin{figure}
	\includegraphics[width=.95\columnwidth]{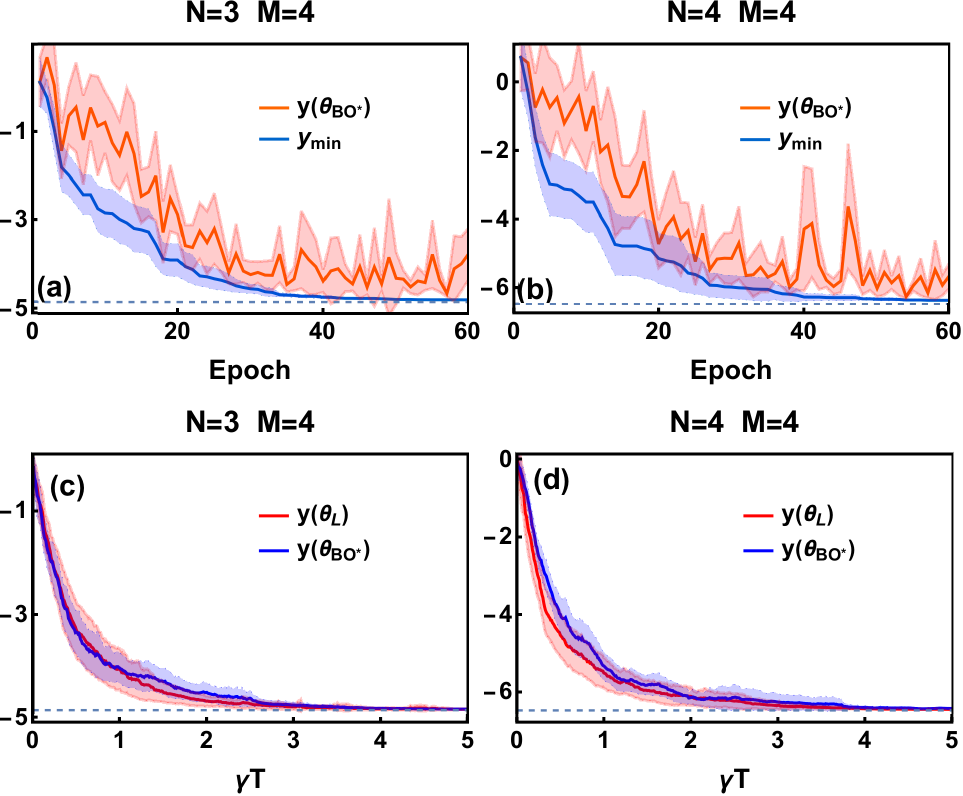}
	\caption{(a),(b) Energy expectation of non-interacting Hamiltonian. Red curves are the mean value of $y(\theta^*_{\text{BO}})$ for each epoch. Blue curves are the mean value of the minimal energy expectation $y_{\text{min}}$ during training. The shades are the minimal energy expectation standard variance of 10 different training processes. (c),(d) Energy expectation during MFCP with 20 different initial states. Red curves are the mean value of energy expectation driven by analytical optimal operators with parameters $\theta_{L}$. Red shades are the energy standard variance. Blue curves are the mean value of energy expectation driven by operators after BO training $\theta_{\text{BO}}$. Gray dashed lines are the ground state energies without interaction. The true ground state energy is calculated as $E_0/J=-4.854$ for $N=3,M=4$ and $E_0/J=-6.472$ for $N=4,M=4$ respectively. } 
	\label{train_non}
\end{figure} 
\textit{Non-interacting case.}  In this part, we apply the BO training process to the MFCP in the non-interaction regime with $U=0$. We aim to search the optimal parameters $\theta^*$ which determine the formalism of the measurement and feedback operators  with limited number of repeated dynamic evolution process. Setting the hopping amplitude $J=1$ as the unit energy and $\gamma=0.1, \delta t=0.1$ and total evolution time $\gamma T=5$ which is long enough for the system to evolve to the target state. During the training process, restricting all the parameters  $\theta\in(-5,5)$. Initially, $n_0$ set of random sample points are provided with $D_0=\{(\theta_i, y_i)|i=1,\cdots,n_0=10)\}$. Due to the long time evolution, the initial state dependence is negligible. Each label $y_i=y(\theta_i)$ is the average of first 25 quantum trajectories with minimal energy expectation over $50$  stochastic dynamic evolutions. During the BO training process, training dataset is extended by candidate optimal datapoint $\{\theta_{\rm BO}^*, y^*\}$ in each iterative loop or epoch as shown in Fig. \ref{flow} (b). 

When $M=4$ and total number of particles $N=3$ with fractional filling, plotting the energy $y(\theta^*_{\rm BO})$ of the candidate datapoint generated by the BO model in Fig. \ref{train_non} (a). As shown by the red dashed line, each line is the average of 10 independent training process. The shadow regime around the red dashed lines stands for the variance of the independent training process. Taking into account of all the data points in the training dataset, the minimal energy $y_{\rm min}$ is plotted by the blue line in Fig. \ref{train_non}(a) where similar average of the training process is applied. It can be found that the variance of the minimal energy $y_{\rm min}$ decreases with large training epochs. The system converges into the ground state energy (the dotted line) with high probability. After training, BO method provides the optimal parameters $\theta^*$ and the associated measurement and feedback operator can drive  the system to the ground state. As shown in Fig. \ref{train_non} (c),  the blue solid line presented the state energy $E_{\rm BO}$ upon average of 20 different initial states.  Analytically, in the non-interaction regime, numerical diagonalization of the semi-positive matrix $\mathcal{K}$ provides two zero eigenvalues. A combination of the two corresponding eigenvectors which guarantees $\beta_1=0$ for the corresponding feedback operators is chosen as the optimal coefficients $\theta_{\rm L}$. The corresponding dynamic evolution $y(\theta_{\rm L})$ is plotted in Fig. \ref{train_non} (c) which provides an analytic benchmark for our BO strategy for the MFCP. The result indicates that the parameters learning by the BO strategy is indeed guide the system to the ground state, as good as the benchmark. For integer filling with $M=N=4$, similar results are also presented in Fig. \ref{train_non} (b) and (d).

\begin{figure}
	\includegraphics[width=.95\columnwidth]{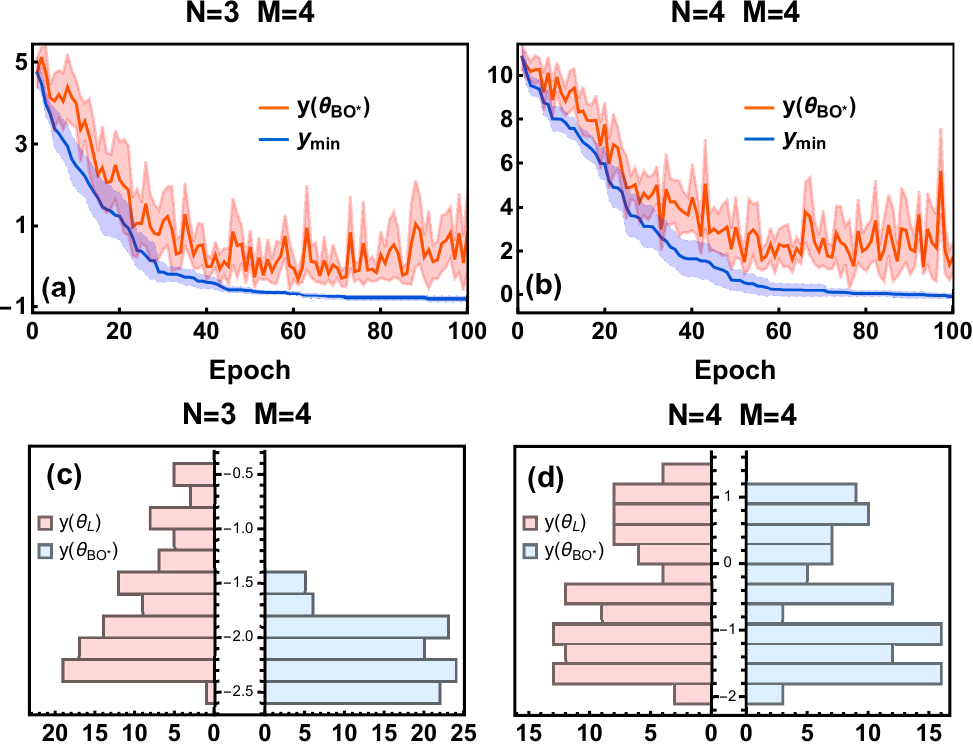}
	\caption{(a),(b) Energy expectation of strong interaction system. Red curves are the mean value of $E_{\rm BO}=y(\theta^*_{\text{BO}})$ for each epoch. Blue curves are the mean value of the minimal energy expectation $E_{min}$ during training. The shades are the minimal energy expectation standard variance of 10 different training processes. With strong interactions,  true ground state energy is calculated as $E_0/J=-2.771$ for $N=3,M=4$ and $E_0/J=-2.204$ for $N=4,M=4$. (c),(d) Histogram of the first 100 small state energy among 200 dynamic trajectories. The light-red bar denotes the MFCP driven by $\theta_{L}$ and the light-blue bar denotes the MFCP driven by $\theta_{\text{BO}}$ } 
	\label{train_inter}
\end{figure}

\textit{Strongly interacting case.} Here we set the interaction strength $U/J=5$. Fig. \ref{train_inter} (a) and (b) show the BO training results for both fractional and integer filling. It can be found that energy expectations $y(\theta^*_{\text{BO}})$ (red lines) have large fluctuations even the minimal energy (blue lines) expectations are converged. Compared to the non-interacting case, it is more harder to reach the ground state energy. In this situation, analytically, the eigenvalues of the matrix $\mathcal{K}$ are all positive. The eigenvector for the minimal eigenvalue is chosen with $\langle \hat{L}^\dagger_{\mathcal{K}}\hat{L}_{\mathcal{K}} \rangle=0.0025$ and $0.001$ for the fractional and integer filling respectively. For the parameters learning by the BO method, the corresponding $\langle \hat{L}^\dagger_{\rm BO}\hat{L}_{\rm BO} \rangle$ is calculated as $0.005$ and $0.0004$ respectively which is the same order of the analytic benchmark result. The deviation of $\langle\hat{L}^\dagger\hat{L}\rangle $ from zero is possibly due to the incomplete  of the operator basis. For practical analysis, provided by the optimal parameters $\theta$, conducting the dynamic evolution for the measurement and feedback process to the final state for $200$ times and plotting the first $100$ trajectories with minimal energy in Fig.\ref{train_inter} (c) and (d). The histogram of the distribution of the state energy indicates a larger occupation of the low energy state. Namely, adopting the parameters provided by the BO method has a larger probability to drive the system to the target state.

\vspace{5pt}

\textbf{Discussion}. 
In this work, we use machine learning method, Bayesian optimization (BO), to optimize the operators for the measurement and feedback process to drive the system to a target state. Taking the one dimensional Bose-Hubbard model as an example, we show that BO can give us a parameter set to drive systems to the low energy state for the given operator basis. Our discussions provide a general scheme for optimizing the controlling parameters, including preparing different classes of target states beyond ground states. Applying this machine learning method only asks for these conditions. First, the entire control process can be quantified. All parameters $\theta$ can be quantitatively determined and can be digitally described. Then once these parameters are fixed, the label $y(\theta)$ is also determined and repeatable. The mapping between parameters and labels can be complicated, and querying labels would be an expensive cost. Our scheme is also model free. We do not need to design a specific machine learning scheme for every optimization task. 

We can consider a number of generalizations of such studies. Firstly, it is interesting to reveal the supremacy of our protocol in quantum simulation and quantum computation, such as infault-tolerant quantum computations \cite{shor1996fault,aharonov1997fault,preskill1998fault,gottesman1998theory}. Then, there still exist many open questions in quantum dynamical phase transition in open systems and some dissipative-driven phase transition \cite{PhysRevA.87.023831,PhysRevB.93.014307,PhysRevA.95.043826}. So MFCP with ML may provide new ways to realize novel forms of dissipative criticality. What's more, recent research indicates that measurements would induce a new kind of phase transition in open system \cite{skinner2019measurement,choi2020quantum,tang2020measurement,fan2021self} , which means that the measurement-feedback control process also has some critical properties.

\vspace{5pt}

\textbf{Acknowledgment.} 
YW is supported by the National Program on Key Basic Research Project of China (Grant No. 2021YFA1400900) and  the National Natural Science Foundation of China (Grant No. 12174236). PZ is partly supported by the Walter Burke Institute for Theoretical Physics at Caltech.  JY is supported by the National Natural Science Foundation of China (Grant No. 11904190).

\bibliographystyle{unsrt}  
\bibliography{BOMFbib}


\end{document}